\documentclass[preprint,showpacs,preprintnumbers,amsmath,amssymb]{revtex4}
\usepackage{epsfig}
\usepackage{graphicx}% Include figure files
\usepackage{dcolumn}% Align table columns on decimal point
\usepackage{bm}% bold math
\usepackage{threeparttable}

\def\beq{\begin{equation}}
\def\eeq{\end{equation}}
\def\eeqn{\end{equation}}
\newcommand\iden{\leavevmode\hbox{\small1\normalsize\kern-.33em1}}

\newcommand{\bea} {\begin{eqnarray}}
\newcommand{\eea} {\end{eqnarray}}

\let\jnfont=\rm
\def\NPB#1,{{\jnfont Nucl.\ Phys.\ B }{\bf #1},}
\def\PLB#1,{{\jnfont Phys.\ Lett.\ B }{\bf #1},}
\def\EPJC#1,{{\jnfont Eur.\ Phys.\ Jour.\ C }{\bf #1},}
\def\PRD#1,{{\jnfont Phys.\ Rev.\ D }{\bf #1},}
\def\PRL#1,{{\jnfont Phys.\ Rev.\ Lett.\ }{\bf #1},}
\def\MPLA#1,{{\jnfont Mod.\ Phys.\ Lett.\ A }{\bf #1},}
\def\JPG#1,{{\jnfont J.\ Phys.\ G }{\bf #1},}
\def\CTP#1,{{\jnfont Commun.\ Theor.\ Phys.\ }{\bf #1},}
\def\JHEP#1,{{\jnfont JHEP \ }{\bf #1},}
\def\NPPS#1,{{\jnfont Nucl.\ Phys.\ Proc.\ Suppl.\ }{\bf #1},}
\def\CPC#1,{{\jnfont Comput.\ Phys.\ Commun.\ }{\bf #1},}
\def\CPL#1,{{\jnfont Chin.\ Phys.\ Lett. }{\bf #1},}
\def\APPB#1,{{\jnfont Acta\ Phys.\ Polon.\ B }{\bf #1},}

\def\lsim{\raise0.3ex\hbox{$<$\kern-0.75em\raise-1.1ex\hbox{$\sim$}}}
\def\gsim{\raise0.3ex\hbox{$>$\kern-0.75em\raise-1.1ex\hbox{$\sim$}}}
\def\PR#1,{{\jnfont Phys.\ Rept. }{\bf #1},}
\def\CHC#1,{{\jnfont Chin.\ Phys.\ C }{\bf #1},}
\begin{document}

\title{\ \\[10mm]
An extension of two-Higgs-doublet model and the excesses of 750 GeV diphoton,
muon g-2 and $h\to\mu\tau$ }
\author{Xiao-Fang Han$^{1}$, Lei Wang$^{2,1}$, Jin Min Yang$^{3,4}$}
 \affiliation{$^1$ Department of Physics, Yantai University, Yantai
264005, P. R. China\\
$^2$ IFIC, Universitat de Val$\grave{e}$ncia-CSIC, Apt. Correus
22085, E-46071 Val$\grave{e}$ncia, Spain\\
$^3$ Institute of Theoretical Physics, Academia Sinica, Beijing 100190, China\\
$^4$ Department of Physics, Tohoku University, Sendai 980-8578, Japan
}

%---------------------------------------------------------------------------

\begin{abstract}
In this paper we simultaneously explain the excesses of the 750 GeV diphoton,
muon g-2 and $h\to \mu\tau$ in an extension of the two-Higgs-doublet model
(2HDM) with additional vector-like fermions and a CP-odd scalar singlet ($P$)
which is identified as the 750 GeV resonance. This 750 GeV  resonance
has a mixing with the CP-odd scalar ($A$) in 2HDM, which leads to a
coupling between $P$ and the SM particles as well as a coupling
between $A$ and the vector-like fermions. Such a mixing and couplings
are strongly constrained by $\tau\to\mu\gamma$, muon g-2 and the 750 GeV
diphoton data. We scan over the parameter space and find that such an extension
can simultaneously account for the observed excesses of 750 GeV diphoton, muon g-2
and $h\to \mu\tau$.
The 750 GeV resonance decays in exotic modes, such as $P\to hA$, $P\to HZ$, $P\to HA$ and
$P\to W^\pm H^\mp$, and its width can be dozens of GeV and is sensitive to the mixing angle.
\end{abstract}
 \pacs{12.60.Fr, 14.80.Ec, 14.80.Bn}

\maketitle

\section{Introduction}
Very recently, the ATLAS and CMS collaborations have both reported an
excess of 750 GeV diphoton resonance \cite{750}, with a local
significance of $3.6\sigma$ and $2.6\sigma$ respectively.
Combining the 8 and 13 TeV data,
the production cross section times the branching ratio is around
4.47$\pm$1.86 fb for CMS and 10.6$\pm$2.9 fb for ATLAS \cite{1512.04939}.
However, there are no excesses for dijet \cite{dijet},
 $t\bar{t}$ \cite{ditt}, diboson or dilepton channels, which gives a challenge
to possible new physics explanations of the 750 GeV diphoton resonance
\cite{1512.04939,work1,work2,work3,work4,work5,work6,work7,work8,work9}.

In addition, the CMS has reported a $2.4\sigma$ excess in
the lepton-flavor-violating (LFV) Higgs decay $h \to \mu\tau$ (here $h$ is
the 125 GeV SM-like Higgs), i.e., $Br(h\to \mu\tau)=(0.84^{+0.39}_{-0.37})\%$
\cite{cmstamu}, while the ATLAS data is $Br(h\to \mu\tau)=(0.7\pm0.62)\%$ \cite{atlastamu}.
This excess can be explained in the general two-Higgs-doublet model (2HDM) with
LFV Higgs interactions. Also such a model can give a sizable
positive contribution to the muon anomalous magnetic moment (muon g-2)
and accommodate the long-standing anomaly \cite{2hdmtamu1,2hdmtamu2,2hdmtamu3}.

Attempting to simultaneously explain the excesses of 750 GeV diphoton, $h\to
\mu\tau$ and muon g-2, we in this work introduce additional vector-like fermions
and a CP-odd scalar singlet ($P$) to the general 2HDM. The singlet
$P$ is identified as the 750 GeV resonance, which has a mixing with
the CP-odd scalar ($A$) in the original 2HDM. Therefore, the model
can lead to the $P$ couplings to SM particles and the $A$ couplings
to vector-like fermions. In addition to the 125 GeV Higgs and 750 GeV
resonance data, the LFV Higgs decay $\tau\to \mu\gamma$ can give strong
constraints on the couplings and mixing. The dominant decays of the
750 GeV resonance can be some exotic modes, such as $P\to h A$, $P\to H
A$, $P\to H Z$ and $P\to W^{\pm} H^{\mp}$. Considering various
relevant experimental constraints, we examine the diphoton
production and decay of the 750 GeV resonance, as well as
muon g-2 and $h\to \mu\tau$.

Our work is organized as follows. In Sec. II we introduce
additional vector-like fermions and a CP-odd scalar singlet to the
2HDM. In Sec. III we perform numerical
calculations and discuss the muon g-2, $h\to \mu\tau$ and the
diphoton production and decay of the 750 GeV resonance in the allowed
parameter space. Finally, we give our conclusion in Sec. IV.

\section{Model}
We introduce a CP-odd scalar singlet field $P_0$ to the general 2HDM
with the assumption that $P_0$ does not develop a vacuum expectation value
(VEV). The Higgs potential is given by \cite{2hdmtamu3,2h-poten}
\beq
\mathrm{V} = \mathrm{V}
_{2HDM}+\frac{1}{2}m^2_{P_0}P_0^2+\frac{\lambda_{P_0}}{4}P_0^4 -
i\mu P_0 \Phi_1^{\dagger}\Phi_2+h.c.,
\eeq
with
\begin{eqnarray}
\label{V2HDM} \mathrm{V}_{2HDM} &=& \mu_1 (\Phi_1^{\dagger} \Phi_1)
+ \mu_2 (\Phi_2^{\dagger}
\Phi_2) + \left[\mu_3 \Phi_1^{\dagger} \Phi_2 + \rm h.c.\right]\nonumber \\
&&+ \lambda_1  (\Phi_1^{\dagger} \Phi_1)^2 +
\lambda_2 (\Phi_2^{\dagger} \Phi_2)^2 + \lambda_3
(\Phi_1^{\dagger} \Phi_1)(\Phi_2^{\dagger} \Phi_2) + \lambda_4
(\Phi_1^{\dagger}
\Phi_2)(\Phi_2^{\dagger} \Phi_1) \nonumber \\
&&+ \left[\lambda_5 (\Phi_1^{\dagger} \Phi_2)^2 + \rm
h.c.\right]+ \left[\lambda_6 (\Phi_1^{\dagger} \Phi_1)
(\Phi_1^{\dagger} \Phi_2) + \rm h.c.\right] \nonumber \\
&& + \left[\lambda_7 (\Phi_2^{\dagger} \Phi_2) (\Phi_1^{\dagger}
\Phi_2) + \rm h.c.\right].
\end{eqnarray}
In the Higgs basis,  the $\Phi_1$ field has a VEV $v=$246 GeV, and
the VEV of $\Phi_2$ field is zero. The two complex scalar doublets
with hypercharge $Y = 1$ can be expressed as
\begin{equation}
\Phi_1=\left(\begin{array}{c} G^+ \\
\frac{1}{\sqrt{2}}\,(v+\rho_1+iG_0)
\end{array}\right)\,, \ \ \
\Phi_2=\left(\begin{array}{c} H^+ \\
\frac{1}{\sqrt{2}}\,(\rho_2+iA_0)
\end{array}\right).
\end{equation}
 The Nambu-Goldstone bosons $G^0$ and $G^+$ are eaten by the gauge bosons.
The physical CP-even Higgs bosons $h$ and $H$ are the linear combinations
of $\rho_1$ and $\rho_2$:
\begin{equation}
\left(\begin{array}{c} \rho_1 \\
\rho_2
\end{array}\right)\, =\ \
\left(\begin{array}{c} ~\cos\alpha~~~~\sin\alpha \\
-\sin\alpha~~~~\cos\alpha
\end{array}\right)\,
\left(\begin{array}{c} h \\
H
\end{array}\right),\,
\end{equation}
where $\tan 2\alpha=2\lambda_6v^2/(m^2_{h22}-m^2_{h11})$ with
\beq
m_{h11}^2=2\lambda_1 v^2,~~~~~~m_{h22}^2=m^2_{H^{\pm}}+ v^2(\frac{1}{2}\lambda_4+\lambda_5).
\eeq
The masses of two CP-even Higgs bosons are given as
\beq
m_{h,H}^2=\frac{1}{2}\left[m^2_{h11}
+m^2_{h22}\mp \sqrt{(m^2_{h11}-m^2_{h22})^2+4\lambda_6^2v^4}\right].
\eeq
The field $H^+$ is the mass eigenstate of the charged Higgs boson,
and the CP-odd Higgs field $A_0$ has a mixing with $P_0$:
\begin{equation}
\left(\begin{array}{c} A_0 \\
P_0
\end{array}\right)\, =\ \
\left(\begin{array}{c} \cos\theta~~~-\sin\theta \\
\sin\theta~~~~~\cos\theta
\end{array}\right)\,
\left(\begin{array}{c} A \\
P
\end{array}\right) ,
\end{equation}
where $\tan2\theta=2\mu v/(m_{A_0}^2-m_{P_0}^2)$ with
\beq
m_{A_0}^2=m^2_{H^{\pm}}+ v^2(\frac{1}{2}\lambda_4-\lambda_5). \eeq
The masses of two CP-odd scalars are given as
\beq
m_{A,P}^2=\frac{1}{2}\left[m_{A_0}^2+m_{P_0}^2\mp \sqrt{(m_{A_0}^2-m_{P_0}^2)^2+4\mu^2v^2}\right].
\eeq
The 750 GeV Higgs boson $P$ couplings to other Higgs bosons and gauge bosons
as
\bea
&&PAh:~c_\theta s_\theta v\left[(\lambda_3 + \lambda_4 -2 \lambda_5) c_\alpha -\lambda_7 s_\alpha\right] -\frac{1}{4v} (m_A^2-m_P^2) s_{4\theta} c_\alpha,\nonumber\\
&&PAH:~c_\theta s_\theta v\left[(\lambda_3 + \lambda_4 -2 \lambda_5) s_\alpha +\lambda_7 c_\alpha\right] -\frac{1}{4v} (m_A^2-m_P^2) s_{4\theta} s_\alpha,\nonumber\\
&&PhZ:~-\frac{e}{2s_Wc_W} s_\alpha s_\theta (p_1 - p_2)^\mu,\nonumber\\
&&PHZ:~\frac{e}{2s_Wc_W} c_\alpha s_\theta (p_1 - p_2)^\mu,\nonumber\\
&&PH^{\pm}W^{\mp}:\frac{e}{2s_W} s_\theta (p_2 - p_1)^\mu.
\eea
The general Yukawa interactions of the SM fermions are given by
 \bea
- {\cal L} &=& y_u\,\overline{Q}_L \,
\tilde{{ \Phi}}_1 \,u_R +\,y_d\,
\overline{Q}_L\,{\Phi}_1 \, d_R\, + \, y_\ell\,\overline{L}_L \, {\Phi}_1\,e_R\nonumber\\
&&+\, \rho^u\,\overline{Q}_L \,
\tilde{{ \Phi}}_2 \,u_R\, +\,\rho^d\,
\overline{Q}_L\,{\Phi}_2 \, d_R\,+\, \rho^\ell \overline{L}_L\, {\Phi}_2\,e_R \,+\, \mbox{h.c.}\,,
\eea
where $Q_L^T=(u_L\,,d_L)$, $L_L^T=(\nu_L\,,l_L)$,
$\widetilde\Phi_{1,2}=i\tau_2 \Phi_{1,2}^*$, and $y_u$, $y_d$, $y_\ell$, $\rho^u$, $\rho^d$
and $\rho^\ell$ are $3 \times 3$ matrices in family space.

Also, we introduce a singlet quark with $\frac{2}{3}$ electric charge
and multiple singlet leptons.
The Yukawa interactions of vector-like fermions are written as
\beq
- {\cal L} = m_T~\bar{T}~ T + i~ y_T~ P_0~ \bar{T}~\gamma_5~ T
 +\sum_i\left(m_{Li}~ \bar{L_i}~ L_i~
+ i~ y_{Li}~ P_0~ \bar{L_i}~\gamma_5~L_i\right).
\eeq

Then we obtain the Yukawa couplings of the neutral Higgs bosons:
\bea
&&
y_{hij}=\frac{m_i^f}{v}c_\alpha\delta_{ij}-\frac{\rho_{ij}^f}{\sqrt{2}}s_\alpha,
~~~y_{Hij}=\frac{m_i^f}{v}s_\alpha\delta_{ij}+\frac{\rho_{ij}^f}{\sqrt{2}}c_\alpha,\nonumber\\
&&y_{Aij}=-i\frac{\rho_{ij}^f}{\sqrt{2}}c_\theta~{\rm (for~u)},~~~y_{Aij}=i \frac{\rho_{ij}^f}{\sqrt{2}}c_\theta~{\rm (for~d,~\ell)},\nonumber\\
&&y_{Pij}=i\frac{\rho_{ij}^f}{\sqrt{2}}s_\theta~{\rm (for~u)},~~~~~y_{Pij}=-i \frac{\rho_{ij}^f}{\sqrt{2}}s_\theta~{\rm (for~d,~\ell)},\nonumber\\
&&y_{ATT}=iy_Ts_\theta,~~~~~~~~~~~~~~~y_{AL_iL_i}=i y_{L_i} s_\theta,\nonumber\\
&&y_{PTT}=iy_Tc_\theta,~~~~~~~~~~~~~~~y_{PL_iL_i}=i y_{L_i} c_\theta.
\eea
For the diagonal matrix elements of $\rho^u$, $\rho^d$ and
$\rho^\ell$, we take
\beq
\rho^u_{ii}=\frac{\sqrt{2}m^u_i}{v}\kappa_u,
~~\rho^d_{ii}=\frac{\sqrt{2}m^d_i}{v}\kappa_d, ~~
\rho^\ell_{ii}=\frac{\sqrt{2}m^\ell_i}{v}\kappa_\ell,
\eeq
which corresponds to the aligned 2HDM \cite{a2hm}. We assume that
$\rho^\ell_{\mu\tau}$ and $\rho^\ell_{\tau\mu}$ are nonzero, and
other nondiagonal matrix elements of $\rho^u$, $\rho^d$ and
$\rho^\ell$ are zero.

The vector-like quark is introduced to make the 750 GeV Higgs singlet
to be produced via the gluon-gluon fusion process. However, the
vector-like quark can also enhance the cross section of $gg\to A$,
which will be constrained by the experimental data from the
ATLAS and CMS searches. Therefore, we expect that the vector-like
leptons play the main role in enhancing the 750 GeV diphoton
production rate. The decay $P\to \gamma\gamma$ can be enhanced by
the vector-like leptons, and its amplitude is proportional to the
couplings and the square of electric charge. Here we do not discuss the
electric charge and coupling of every vector-like lepton as well as
the quantity of vector-like leptons in detail; instead we focus on the
total contribution of vector-like leptons, which depends on
\beq
Y_L=\sum_i y_{Li} Q^2_{Li},
\eeq
where $L_i$ denotes the $i$-th vector-like lepton.

\section{Numerical calculations and discussions}
\subsection{Numerical calculations}
In our calculations, we scan over the parameters in the following range
\bea
&&-0.06<s_\alpha<0.06, ~~~~~~~ -0.3<s_\theta<0.3,\nonumber\\
&&0.05<\rho_{\mu\tau}=\rho_{\tau\mu}<1,~~~-50<\kappa_\ell<50,\nonumber\\
&&0<Y_L<50,~~~~~~~~~~~~~~~~~~~0<\lambda_3,~\lambda_7<4\pi,\nonumber\\
&&200~{\rm GeV}< m_H < 450~{\rm GeV},
\eea
and fix
\bea
&&m_h=125.5~ {\rm GeV} ~~~~ m_P=750~ {\rm GeV},~~~~m_{H^\pm}=m_A=500~ {\rm GeV},\nonumber\\
&&m_{Li}=400~ {\rm GeV},~~~~m_T=700~ {\rm GeV},~~~~y_T=2.0,\nonumber\\
&&\kappa_u=\kappa_d=0.
\eea
During the scan, we consider the following experimental constraints and observables:
\begin{itemize}
\item[(1)] Precision electroweak data. According to the expressions for
the oblique parameters $S$, $T$ and $U$ in the 2HDM \cite{stu}, for
$-0.06<s_\alpha<0.06$ and $c_\alpha\simeq1$, the expressions in
this model are approximately given as
\bea
S&=&\frac{1}{\pi m_Z^2}
\left[ c_\alpha^2 c_\theta^2 F_S(m_Z^2,m_H^2,m_A^2) + c_\alpha^2
s_\theta^2 F_S(m_Z^2,m_H^2,m_P^2)
-F_S(m_Z^2,m_{H^{\pm}}^2,m_{H^{\pm}}^2) \right], \nonumber \\
T&=&\frac{1}{16\pi m_W^2 s_W^2} \left[ -c_\alpha^2 c_\theta^2 F_T(m_H^2,m_A^2) - c_\alpha^2 s_\theta^2 F_T(m_H^2,m_P^2)
+ c_\alpha^2 F_T(m_{H^{\pm}}^2,m_H^2) \right. \nonumber \\
&&+ \left. c_\theta^2 F_T(m_{H^{\pm}}^2,m_A^2) + s_\theta^2 F_T(m_{H^{\pm}}^2,m_P^2)\right], \nonumber\\
U&=&\frac{1}{\pi m_W^2} \left[ c_\alpha^2 F_S(m_W^2,m_{H^{\pm}}^2,m_H^2) -2 F_S(m_W^2,m_{H^{\pm}}^2,m_{H^{\pm}}^2) \right.\nonumber\\
&& \left.+c_\theta^2 F_S(m_W^2,m_{H^{\pm}}^2,m_A^2) + s_\theta^2 F_S(m_W^2,m_{H^{\pm}}^2,m_P^2) \right]\nonumber\\
&&-\frac{1}{\pi m_Z^2} \left[ c_\alpha^2 c_\theta^2 F_S(m_Z^2,m_H^2,m_A^2) + c_\alpha^2 s_\theta^2 F_S(m_Z^2,m_H^2,m_P^2) \right.\nonumber\\
&&\left.- F_S(m_Z^2,m_{H^{\pm}}^2,m_{H^{\pm}}^2) \right],
\eea
where
\beq
F_T(a,b)=\frac{1}{2}(a+b)-\frac{ab}{a-b}\log(\frac{a}{b}),~~F_S(a,b,c)=B_{22}(a,b,c)-B_{22}(0,b,c)
\eeq
with
\bea
&&B_{22}(a,b,c)=\frac{1}{4}\left[b+c-\frac{1}{3}a\right] - \frac{1}{2}\int^1_0 dx~X\log(X-i\epsilon),\nonumber\\
&&
X=bx+c(1-x)-ax(1-x).
\eea
Here we require \cite{pdg2014}
\beq
S=-0.03\pm0.1,~~~T=0.01\pm0.12,~~~U=0.05\pm 0.1
\eeq

\item[(2)] The 125 GeV Higgs data. For $-0.06<s_\alpha<0.06$,
$\kappa_u=\kappa_d=0$ and $-50<\kappa_\ell<50$, the 125 GeV Higgs
couplings to the gauge bosons, up-type quark and down-type quark are
very close to the SM values, but the coupling to $\bar{\tau}\tau$ can have
a sizable deviation from the SM value. The signal strength of
$\bar{\tau}\tau$ channel is
$\hat{\mu}_{\tau\tau}=1.41^{+0.4}_{-0.35}$ from ATLAS \cite{atlastt}
and $\hat{\mu}_{\tau\tau}=0.89^{+0.31}_{-0.28}$ from CMS
\cite{cmstt}. We require $0.33 <\hat{\mu}_{\tau\tau}< 2.21$
and such a bound will give strong constrains on $s_\alpha$ and
$\kappa_\ell$ for which the absolute value of the coupling of the
125 GeV Higgs and $\bar{\tau}\tau$ is around the SM value.

\item[(3)] Non-observation of additional Higgs bosons. For
$-0.06<s_\alpha<0.06$ and $\kappa_u=\kappa_d=0$, the cross sections of
$H$ and $H^{\pm}$ at the collider are very small, and hence $H$ and
$H^{\pm}$ can be hardly constrained by the current experimental data
from the ATLAS and CMS searches. The pseudoscalar $A$ can be
produced via the gluon-gluon fusion process with vector-like quark
loop, and the decay $A\to \gamma\gamma$, $A\to \gamma Z$ and $A\to
ZZ$ can be enhanced by the vector-like quark and leptons at one-loop
level. For $m_A=$500 GeV, we impose the following relevant bounds
at the 8 TeV LHC \cite{8rr,8zz,8zr,8hz,8tata}
\beq
R_{\gamma\gamma} < 6~ {\rm fb},~~~R_{ZZ} < 45~ {\rm
fb},~~~R_{Z\gamma} < 6.8~ {\rm fb},~~~ R_{hZ}~ < 60~ {\rm
fb},~~~R_{\bar{\tau}\tau} < 26~ {\rm fb}.
\eeq

\item[(4)] The 750 GeV resonance data. The 750 GeV Higgs singlet $P$ can be
produced via the gluon-gluon fusion process with vector-like quark
loop, and the decays $A\to \gamma\gamma$, $A\to \gamma Z$ and $A\to
ZZ$ can be enhanced by the vector-like quark and leptons at one-loop
level. Due to the mixing with $A$, the 750 GeV singlet can decay into
the SM particles, such as $hZ$ and $\bar{\tau}\tau$. For the 750 GeV
Higgs singlet, we impose the following bounds at the 8 TeV LHC
\cite{8rr,8zz,8zr,8hz,8tata}
\beq
R_{\gamma\gamma} < 2~ {\rm fb},~~~R_{ZZ} < 12~ {\rm
fb},~~~R_{Z\gamma} < 4~ {\rm fb},~~~ R_{hZ}~ < 19~ {\rm
fb},~~~R_{\bar{\tau}\tau} < 12~ {\rm fb}.
\eeq
At the 13 TeV LHC, we require the 750 GeV diphoton production rate as
\beq
2~{\rm fb}<R_{\gamma\gamma}<10~{\rm fb}.
\eeq

\item[(5)] The data of Br$(h\to \mu\tau)$.
The branching ratio of $h\to \mu\tau$ is given by
\beq
Br(h\to \mu\tau)=\frac{s_\alpha^2 (\rho_{\mu\tau}^2+\rho_{\tau\mu}^2)m_h}{16\pi\Gamma_h},
\eeq
where $\Gamma_h$ is the total width of 125 GeV Higgs. To explain the $h\to \mu\tau$ excess
reported by CMS within $2\sigma$ range, we require
\beq
0.1\%<Br(h\to \mu\tau)<1.62\%.
\eeq

\item[(6)] The muon g-2 data. The dominant contributions to the muon g-2
are from the one-loop diagrams
 with the Higgs LFV coupling \cite{mua1loop},
\bea
  \delta a_{\mu1}&=&\frac{m_\mu m_\tau \rho_{\mu\tau}\rho_{\tau\mu}}{16\pi^2}
  \left[\frac{s^2_{\alpha}(\log\frac{m_h^2}{m_\tau^2}-\frac{3}{2})}{m_h^2}
    +\frac{c^2_{\alpha}(\log\frac{m_H^2}{m_\tau^2}-\frac{3}{2})}{m_H^2} \right.\nonumber\\
&&\left.~~~~~~~~~~~~~~~~~~~~~~
-\frac{c^2_{\theta}\log(\frac{m_A^2}{m_\tau^2}-\frac{3}{2})}{m_A^2}
-\frac{s^2_{\theta}\log(\frac{m_P^2}{m_\tau^2}-\frac{3}{2})}{m_P^2}
\right].
  \label{mua1}
\eea
The muon g-2 can be also corrected by the two-loop Barr-Zee diagrams
with the fermions loops, $W$ and Goldstone loops. Using the well-known
classical formulates \cite{mua2loop}, the main contributions
of two-loop Barr-Zee diagrams in this model are given as
\bea \label{mua2}
\delta a_{\mu2} &=&-\frac{\alpha
m_\mu}{4\pi^3m_f}\sum_{\phi=h,H,A,P;f=t,b,\tau,T,L_i} N_f^c~Q_f^2~
y_{\phi\mu\mu}~ y_{\phi ff}~ F_\phi(x_{f\phi})
\nonumber\\
&&+\frac{\alpha m_\mu}{8\pi^3v}\sum_{\phi=h,H} y_{\phi\mu\mu}~g_{\phi WW} \left[3F_H\left(x_{W\phi}\right)
    +\frac{23}{4} F_A\left(x_{W\phi}\right)
     \right.\nonumber\\
    &&\left.
+\frac{3}{4} G\left(x_{W\phi}\right)
+\frac{m_\phi^2}{2 m_W^2}\left\{
    F_H\left(x_{W\phi}\right)-F_A\left(x_{W\phi}\right)
    \right\}\right],
\eea
where $x_{f\phi}=m_{f}^2/m_\phi^2$, $x_{W\phi}=m_W^2/m_\phi^2$,
$g_{H WW}=s_{\alpha}$, $g_{h WW}=c_{\alpha}$ and
\begin{align}
  F_\phi(y)&=F_{H}(y)=\frac{y}{2}\int_0^1 dx \frac{1-2x(1-x)}{x(1-x)-y}\log \frac{x(1-x)}{y}
~~({\rm for}~\phi=h,~H) \\
  F_\phi(y)&=F_A(y) =\frac{y}{2}\int_0^1 dx \frac{1}{x(1-x)-y}\log \frac{x(1-x)}{y}~~
({\rm for}~\phi=A,~P) \\
  G(y)&=-\frac{y}{2}\int_0^1 dx \frac{1}{x(1-x)-y}\left[
    1-\frac{y}{x(1-x)-y}\log \frac{x(1-x)}{y}
    \right].
\end{align}
 The experimental value of muon g-2 excess is \cite{muaexp}
\beq
\delta a_{\mu} =(26.2\pm8.5) \times 10^{-10}.
\eeq

\item[(7)] The data of Br$(\tau\to \mu\gamma)$.
The LFV coupling of the Higgs boson gives the dominant contributions
to the decay $\tau\to\mu\gamma$. The branching ratio of $\tau\to \mu\gamma$ is given by
\beq
\frac{{\rm BR}(\tau \rightarrow \mu \gamma)}{{\rm BR}(\tau \rightarrow \mu \bar{\nu}_\mu \nu_\tau)}
  =\frac{48\pi^3\alpha\left(|A_{1L0}+A_{1Lc}+A_{2L}|^2+|A_{1R0}+A_{1Rc}+A_{2R}|^2\right)}
  {G_F^2},
\eeq
where $A_{1L0}$, $A_{1Lc}$, $A_{1R0}$ and $A_{1Rc}$ are from the one-loop diagrams with the
Higgs bosons and tau lepton \cite{2hdmtamu2}, and
\bea
A_{1L0}&=&\sum_{\phi=h,~H,~A,~P}\frac{y^{*}_{\phi\;\tau \mu}}{16\pi^2 m_\phi^2}
  \left[y^{*}_{\phi\;\tau\tau}\left(\log\frac{m_\phi^2}{m_\tau^2}-\frac{3}{2}\right)
    +\frac{y_{\phi\;\tau\tau}}{6}\right], \\
A_{1Lc}&=&-\frac{(\rho^{e\dagger} \rho^e)^{\mu \tau}}{192\pi^2 m_{H^-}^2}, \\
A_{1R0}&=&A_{1L0}\left({y^{*}_{\phi\; \tau\mu}\rightarrow y_{\phi\; \mu\tau},
    ~~y_{\phi \;\tau \tau} \leftrightarrow y^{*}_{\phi\; \tau\tau}}\right),\\
A_{1Rc}&=&0.
\eea
Here $A_{2L}$ and $A_{2R}$ are from the two-loop Barr-Zee diagrams with the third-generation
fermions loops, vector-like fermions loops and $W$ loops \cite{2hdmtamu2}:
\bea
  A_{2L}&=&-\sum_{\phi=h,H,A,P;f=t,b,\tau,T,L_i}
  \frac{N_C Q_f \alpha}{8\pi^3}
  \frac{y^{*}_{\phi\;\tau\mu}}{m_\tau m_{f}}
  \left[
  Q_f\left\{
    {\rm Re} (y_{\phi\; ff})
    F_H\left(x_{f\phi}\right)
  - i {\rm Im} (y_{\phi\; ff})
  F_A\left(x_{f\phi}\right)\right\}\right.
  \nonumber \\
  &&\left. +\frac{(1-4 s_W^2)(2T_{3f}-4Q_f s_W^2)}{16s_W^2 c_W^2}
  \left\{
    {\rm Re} (y_{\phi\; ff})
    \tilde{F}_H\left(x_{f\phi},x_{fZ}\right)
  - i {\rm Im} (y_{\phi\; ff})
  \tilde{F}_A\left(x_{f\phi},x_{f Z}\right)\right\}\right]
  \nonumber \\
  &&+\sum_{\phi=h,H}\frac{\alpha}{16\pi^3}\frac{g_{\phi WW} y^{*}_{\phi\;\tau\mu}}{m_\tau v}
  \left[
    3F_H\left(x_{W\phi}\right)
    +\frac{23}{4} F_A\left(x_{W\phi}\right)
    \nonumber \right.\\
  &&+\frac{3}{4} G\left(x_{W\phi}\right)
    +\frac{m_\phi^2}{2 m_W^2}\left\{
    F_H\left(x_{W\phi}\right)-F_A\left(x_{W\phi}\right)
    \right\}\nonumber \\
  &&+\frac{1-4s_W^2}{8s_W^2}\left\{\left(
    5-t_W^2+\frac{1-t_W^2}{2 x_{W\phi}}\right) \tilde{F}_H (x_{W\phi}, x_{WZ})
    \right.\nonumber \\
  &&\left.\left.+\left(7-3t_W^2 -\frac{1-t_W^2}{2x_{W\phi}}\right) \tilde{F}_A(x_{W\phi}, x_{WZ})
    +\frac{3}{2}\left\{F_A(x_{W\phi})+G(x_{W\phi})\right\} \right\}   \right],  \label{Barr-Zee} \\
  A_{2R} &=&A_{2L}\left(y^{*}_{\phi\;\tau\mu}\rightarrow
  y_{\phi\;\mu\tau},~i\rightarrow -i\right),
\eea
where $T_{3f}$ denotes the isospin of the fermion,
$t_W^2=s_W^2/c_W^2$, $x_{fZ}=m_{f}^2/m_Z^2$ and $x_{WZ}=m_W^2/m_Z^2$,
and
\begin{align}
  \tilde{F}_H(x,y)&=\frac{xF_H(y)-yF_H(x)}{x-y},\\
  \tilde{F}_A(x,y)&=\frac{xF_A(y)-yF_A(x)}{x-y}.
\end{align}
The terms in the first two lines of Eq.~(\ref{Barr-Zee})
come from the effective $\phi \gamma\gamma$ vertex and $\phi Z\gamma$ vertex
induced by the third-generation fermion loop and vector-like fermion loop.
Other terms are from the effective $\phi \gamma\gamma$ vertex and
$\phi Z\gamma$ vertex induced by the W-boson loop. The current upper bound
of $Br(\tau\to\mu\gamma)$ is \cite{tamurexp1,tamurexp2}
\beq
Br(\tau\to\mu\gamma) < 4.4\times 10^{-8}.
\eeq
\end{itemize}

\subsection{Results and discussions}
%%%%%%%%%%%%%%%%%%%%%
\begin{figure}[tb]
%\begin{center}
 \epsfig{file=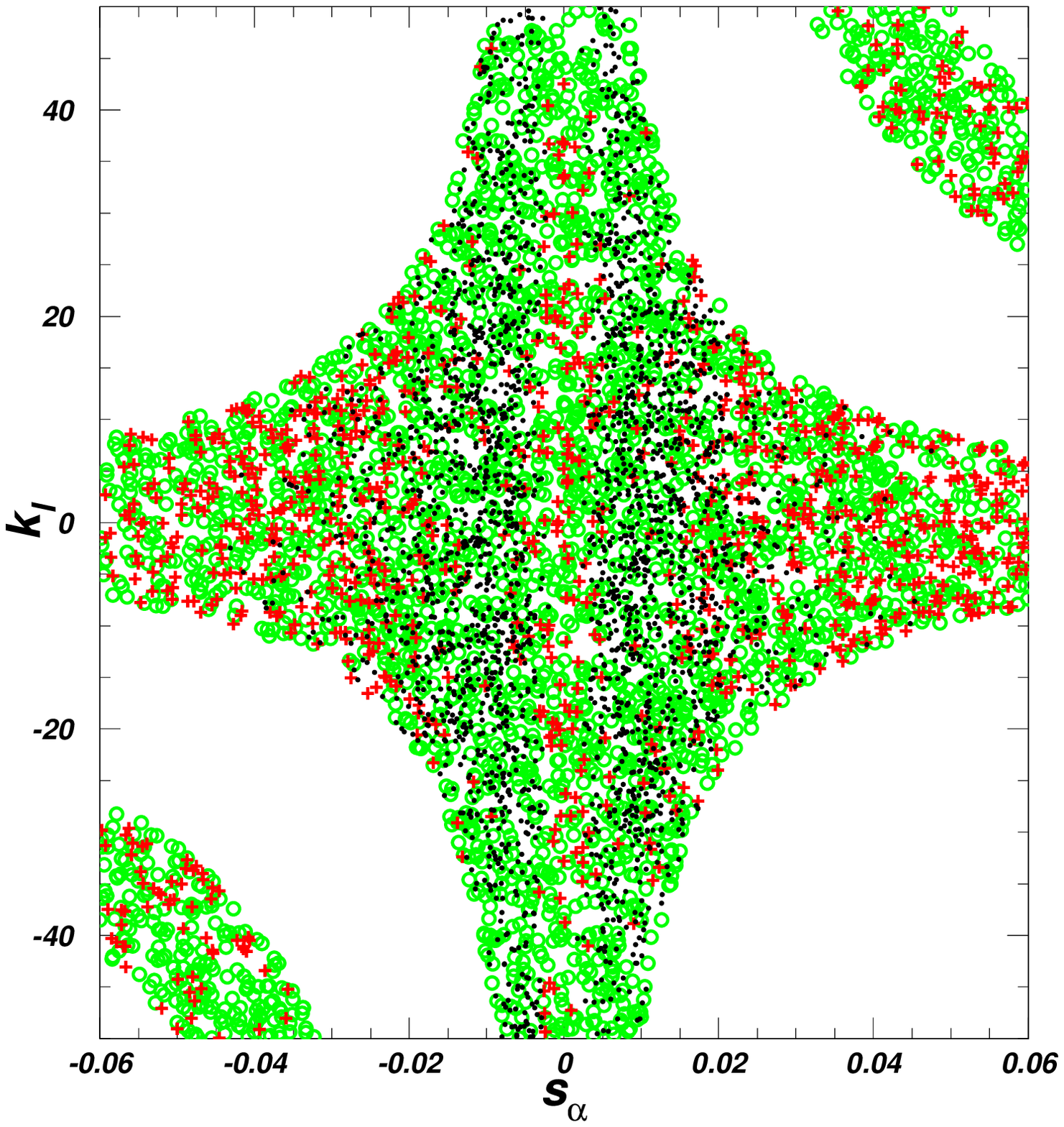,height=7.0cm}
  \epsfig{file=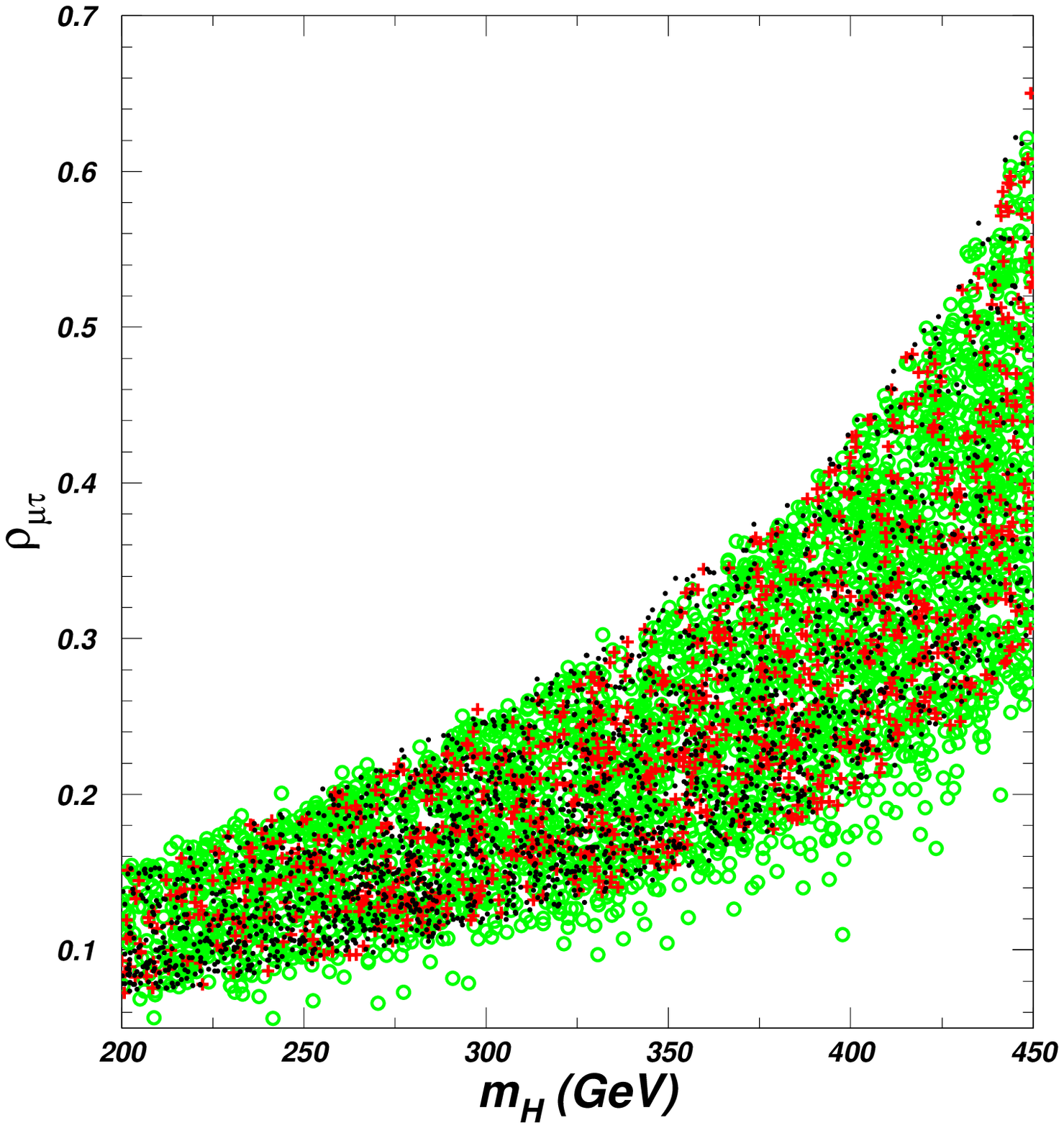,height=7.0cm}
 \epsfig{file=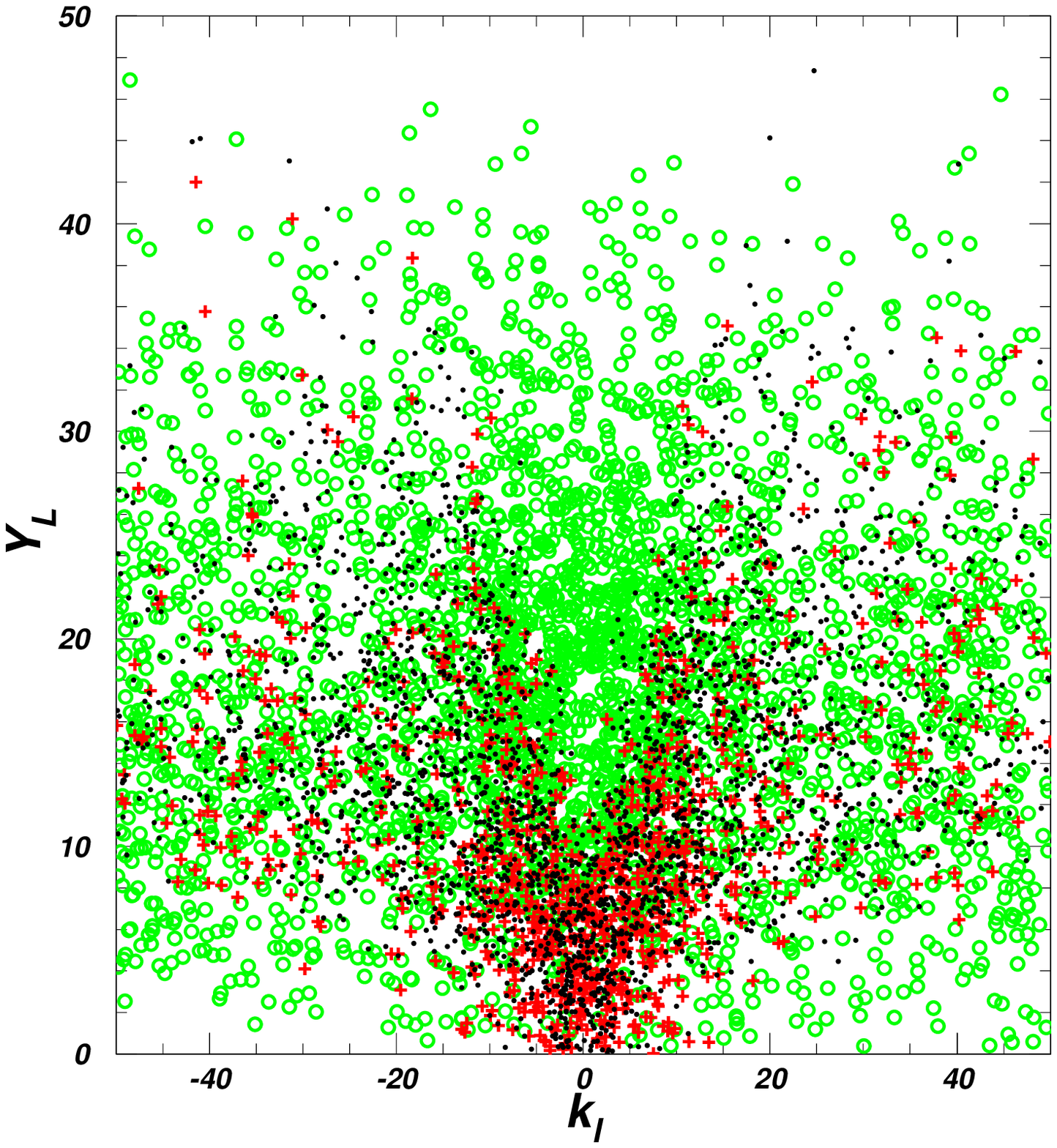,height=7.1cm}
  \epsfig{file=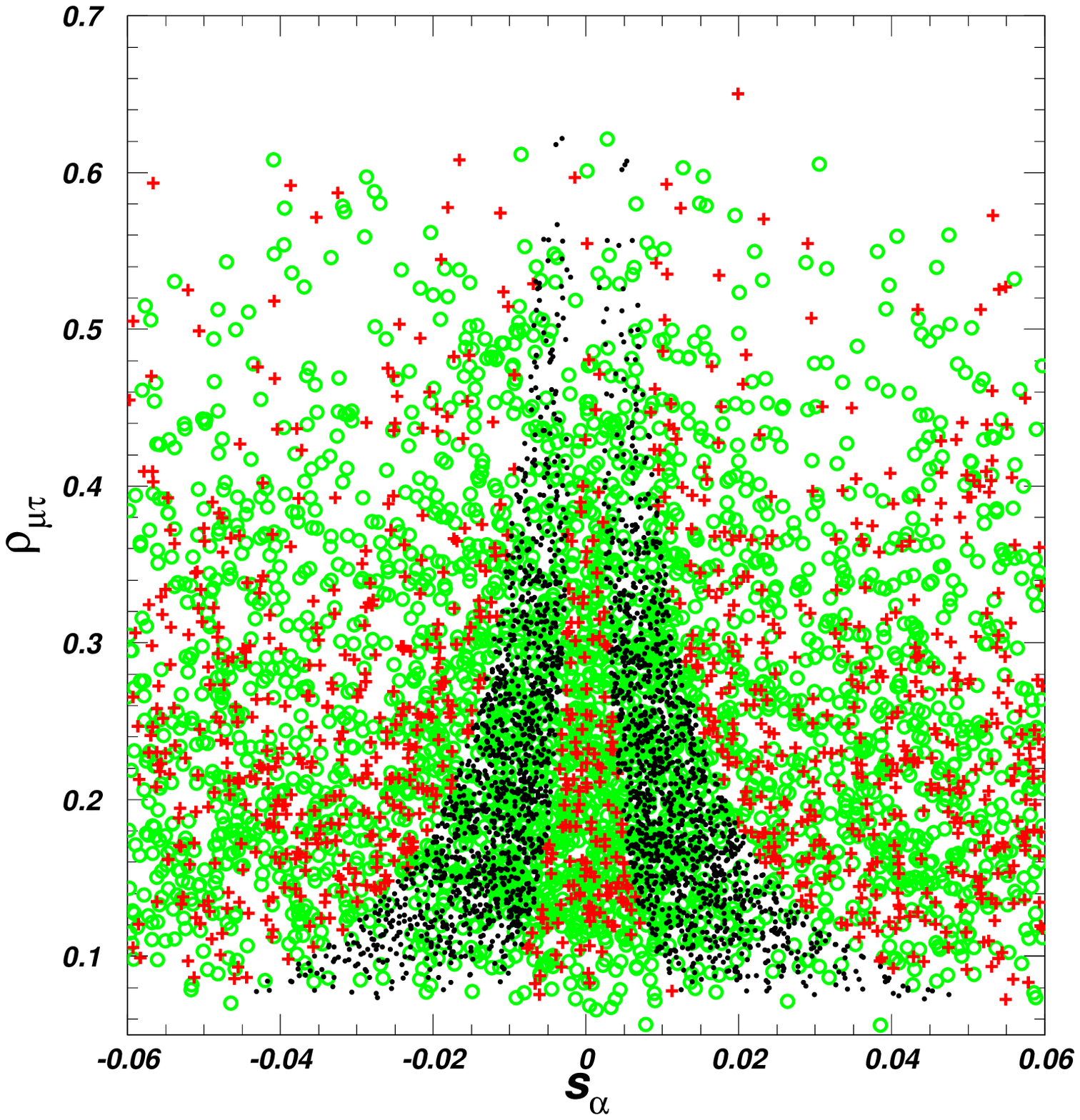,height=7.0cm}
 %\end{center}
\vspace{-0.5cm} \caption{Under the constrains of the oblique parameters and the LHC Higgs data,
the surviving samples projected on the planes of $\kappa_\ell$ versus $s_\alpha$,
 $\rho_{\mu\tau}$ versus $m_H$
, $Y_L$ versus $\kappa_\ell$ and $\rho_{\mu\tau}$ versus $s_\alpha$.
The circles (green) are allowed by the muon g-2,
the pluses (red) allowed by the muon g-2 and $Br(\tau\to\mu\gamma)$,
and the bullets (black) allowed by the muon g-2, $Br(\tau\to\mu\gamma)$ and $Br(h\to\mu\tau)$.}
\label{tamur}
\end{figure}
%%%%%%%%%%%%%%%%%%%%

%%%%%%%%%%%%%%%%%%%%%
\begin{figure}[tb]
%\begin{center}
 \epsfig{file=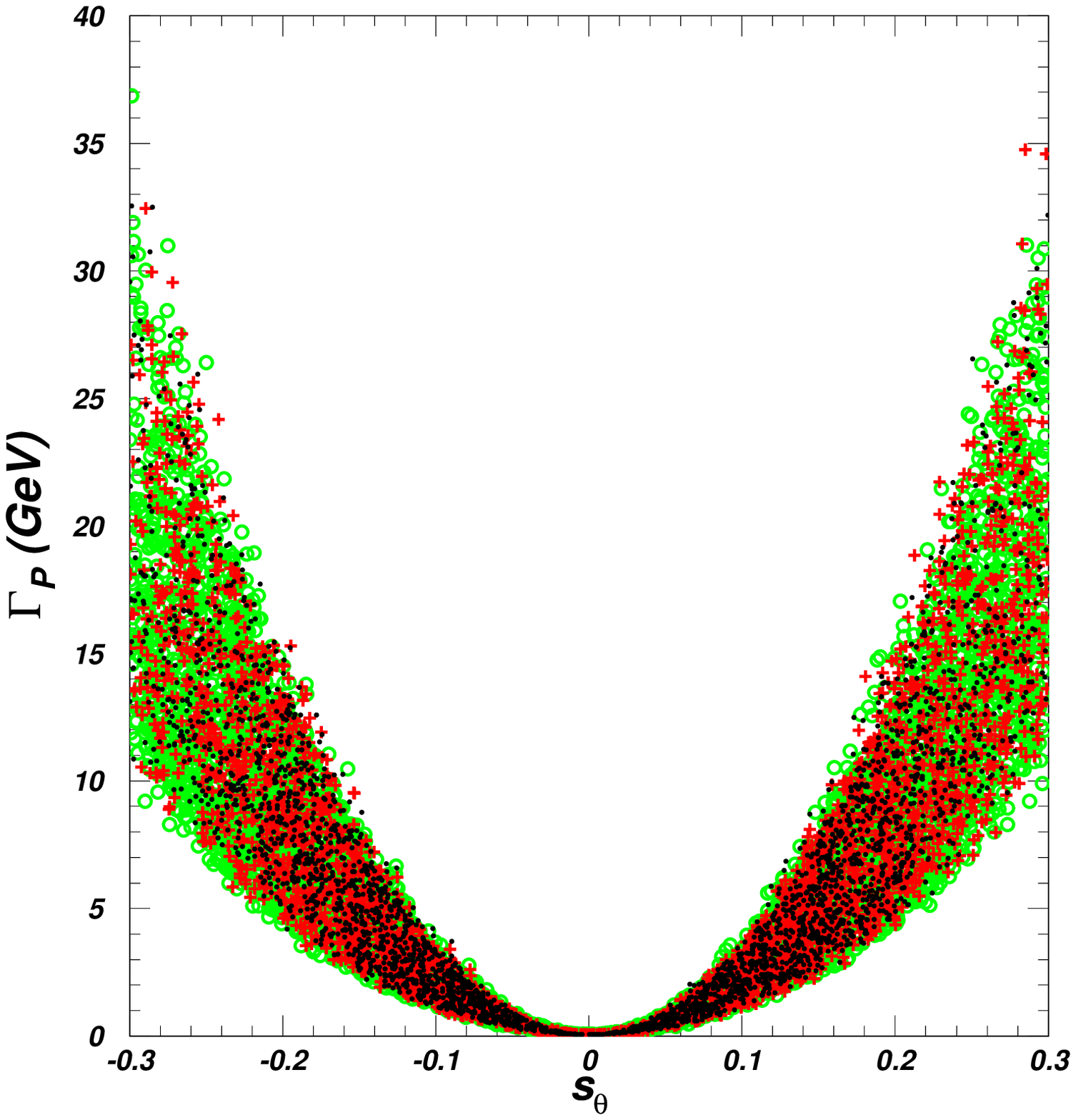,height=5.65cm}
  \epsfig{file=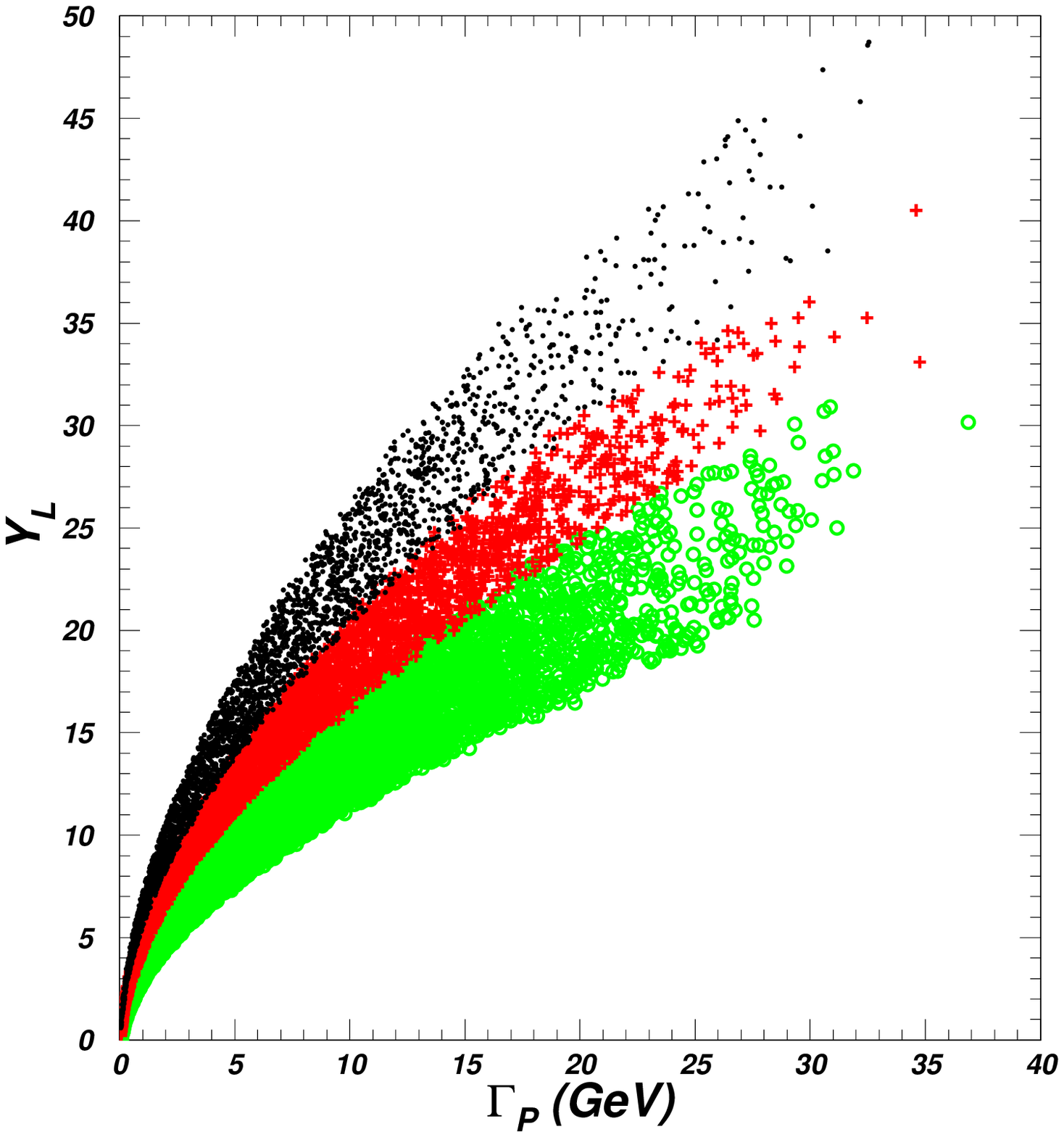,height=5.65cm}
   \epsfig{file=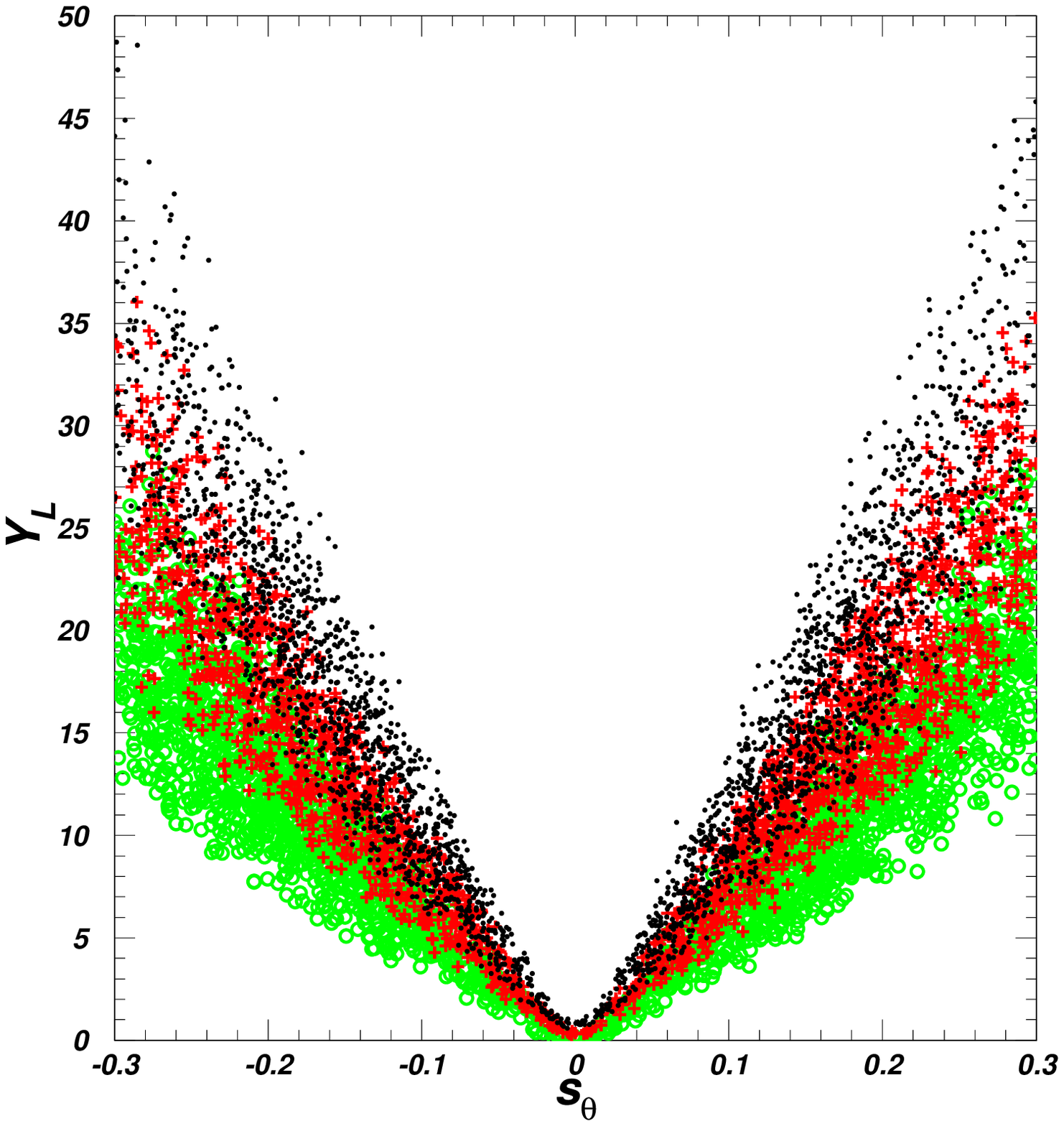,height=5.6cm}
\vspace{-0.5cm}
\caption{Under the constrains of the oblique parameters of electroweak data,
the LHC Higgs data, muon g-2, $BR(\tau\to\mu\gamma)$
and $Br(h\to\mu\tau)$, the surviving samples projected on the planes
of $\Gamma_P$ versus $s_\alpha$, $Y_L$ versus $\Gamma_P$
and $Y_L$ versus $s_\theta$.
Here 2 fb $<R_{\gamma\gamma}<$ 4 fb for the circles (green), 4 fb
$<R_{\gamma\gamma}<$ 6 fb for the pluses (red), and 6 fb
$<R_{\gamma\gamma}<$ 10 fb for the bullets (black), with $R_{\gamma\gamma}$
denoting the 750 GeV Higgs production rate at the 13 TeV LHC.} \label{sigma}
\end{figure}
%%%%%%%%%%%%%%%%%%%%

%%%%%%%%%%%%%%%%%%%%%
\begin{figure}[tb]
%\begin{center}
 \epsfig{file=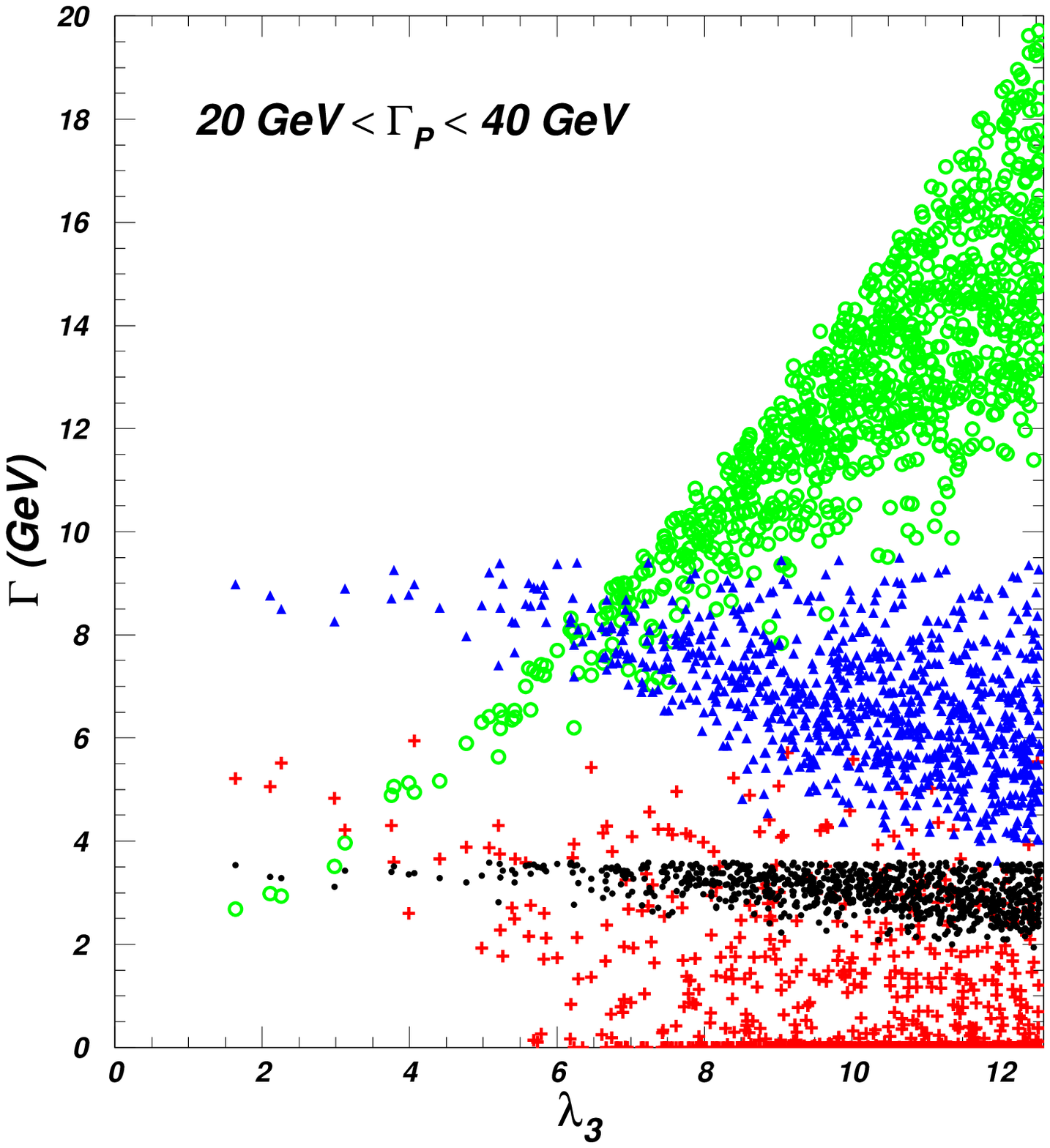,height=5.85cm}
  \epsfig{file=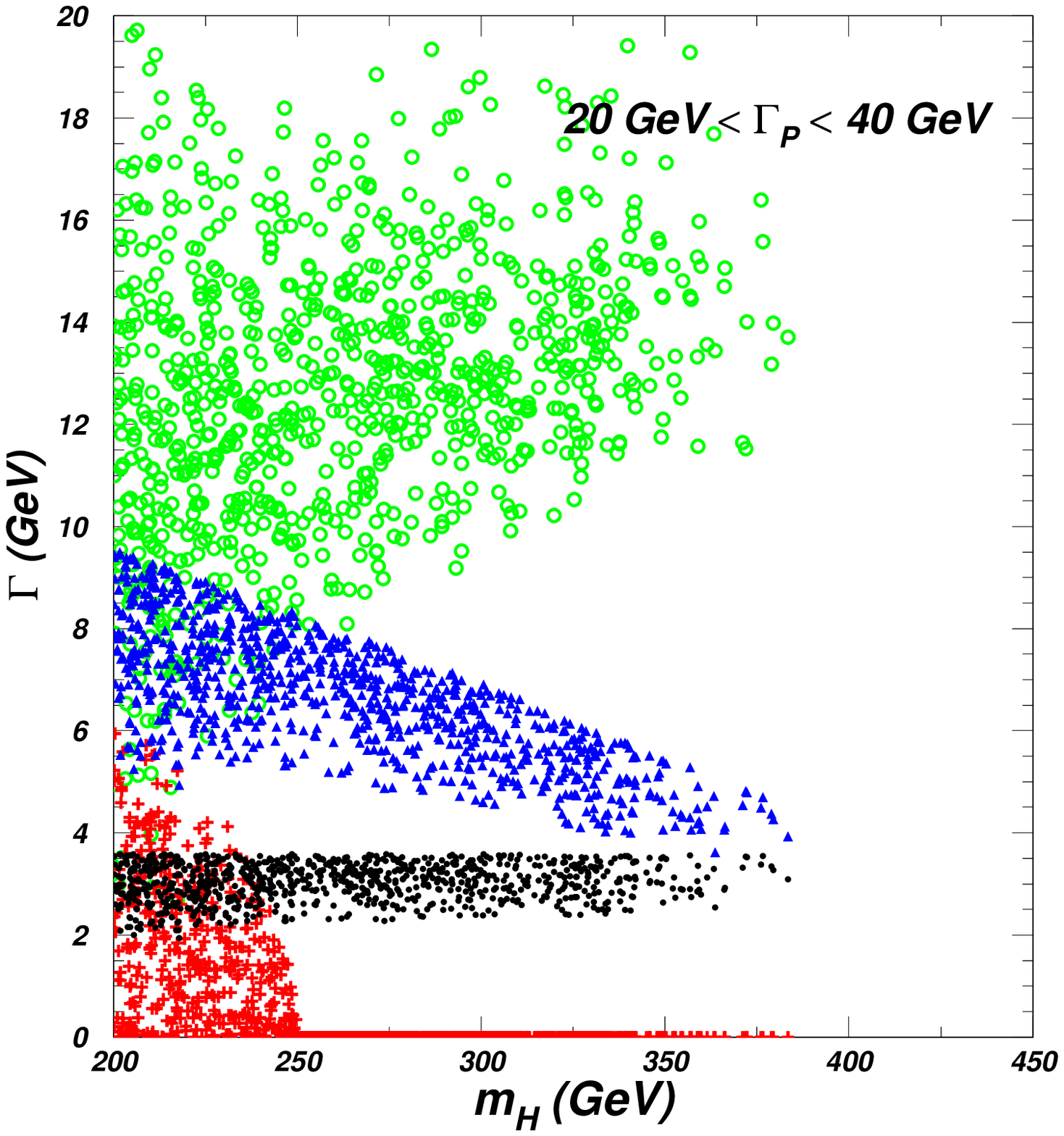,height=5.85cm}
   \epsfig{file=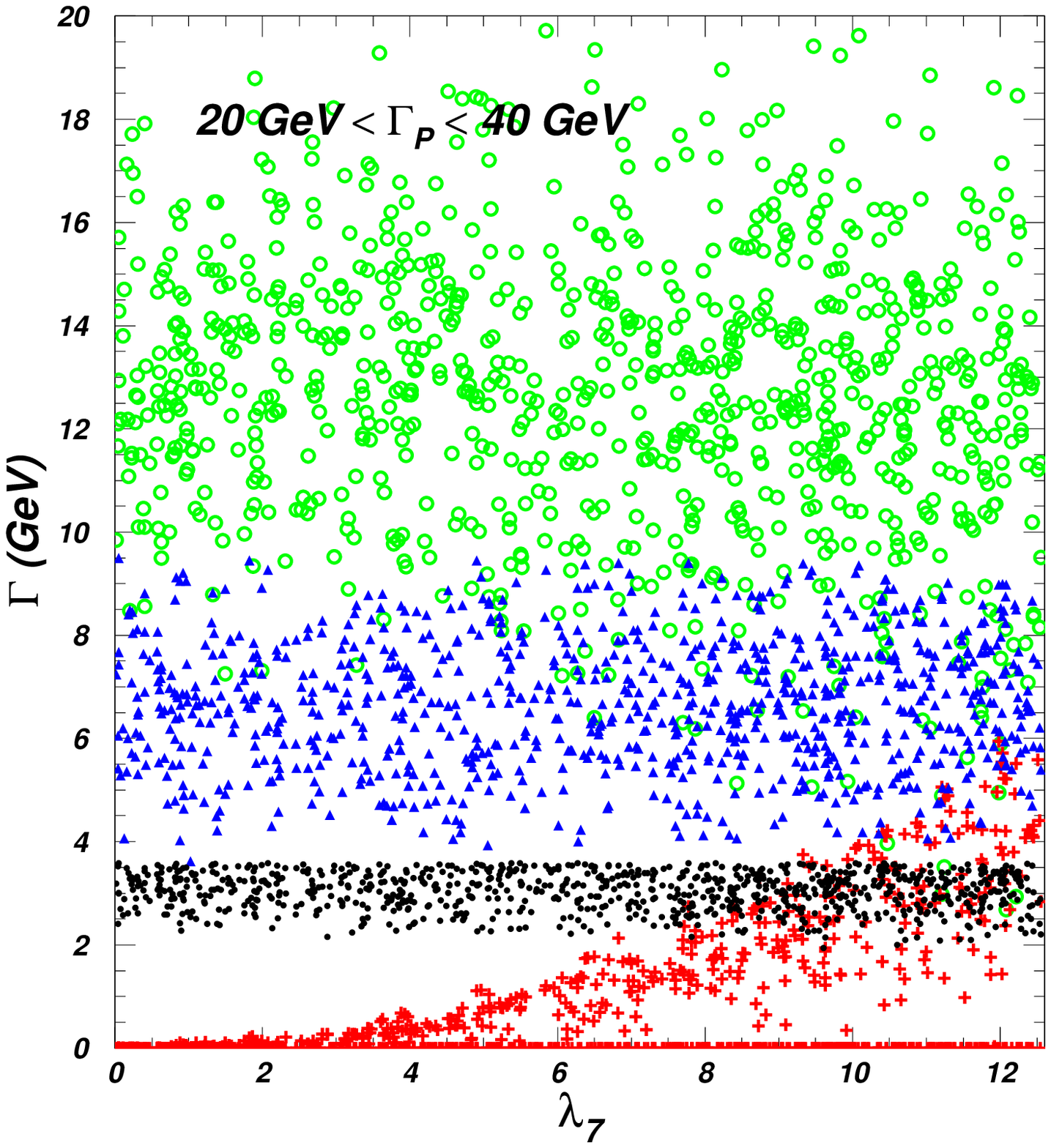,height=5.85cm}
\vspace{-0.5cm} \caption{Under the constrains of the oblique
parameters of electroweak data, the LHC Higgs data, muon g-2, $Br(\tau\to\mu\gamma)$ and
$Br(h\to\mu\tau)$, the surviving samples with 20 GeV $<\Gamma_P<$ 40
GeV projected on the planes of the widths of the main decay modes of
the 750 GeV Higgs versus $\lambda_s$, $m_H$ and $\lambda_7$.
Here $\Gamma(P\to hA)$ for the circles (green), $\Gamma(P\to HZ)$ for the
triangles (blue), $\Gamma(P\to HA)$ for the pluses (red) and
$\Gamma(P\to W^{\pm}H^{\mp})$ for the bullets (black).}
\label{width}
\end{figure}
%%%%%%%%%%%%%%%%%%%%

In Fig. \ref{tamur}, we project the surviving samples on the planes
of $\kappa_\ell$ versus $s_\alpha$, $\rho_{\mu\tau}$ versus $m_H$,
$Y_L$ versus $\kappa_\ell$, and $\rho_{\mu\tau}$ versus $s_\alpha$.
The upper-left panel shows that there is a strong correlation
between $s_\alpha$ and $\kappa_\ell$ due to the experimental
constraints of the $\bar{\tau}\tau$ channel data of 125 GeV Higgs.
The surviving samples have two different 125 GeV Higgs
couplings to $\bar{\tau}\tau$, and their absolute values are around
the SM value. One is the SM-like Higgs coupling with the same sign as the
coupling of the gauge boson, and the other is the Yukawa
coupling with the opposite sign to the coupling of the gauge boson
for a relative large $\kappa_\ell$.

From the upper-right panel of Fig. \ref{tamur}, we see that the muon g-2 favors
$\rho_{\mu\tau}$ to increase with $m_H$.
As shown in Eq. (\ref{mua1}), the muon g-2 can obtain positive
contributions from the $H$ loop and negative contributions from $A$
and $P$ loops for $\rho_{\mu\tau}=\rho_{\tau\mu}$. With the
decreasing of the mass splitting of $H$ and $A$, the cancelation
between the contributions of $H$ and $A$ loops becomes sizable so
that a large $\rho_{\mu\tau}$ is required to enhance the muon g-2.
From the lower-left panel, we see that the upper bound of $\tau\to
\mu\gamma$ favors a large absolute value of $\kappa_\ell$ for a
large $Y_L$. The vector-like leptons with a large $Y_L$ can
sizably enhance $Br(\tau\to\mu\gamma)$ via the two-loop Barr-Zee
diagrams, and such contributions can be partially canceled by the
one-loop diagram for a properly large $\kappa_\ell$ and a proper sign of
$s_\theta$. The lower-right panel shows that the experimental data
of $Br(h\to \mu\tau)$ requires $\rho_{\mu\tau}$ to increase with
the decreasing of the absolute value of $s_\alpha$, and -0.05
$<s_\alpha<$ 0.05 and 0.05 $<\rho_{\mu\tau}<$ 0.7 are favored by the
$Br(h\to\mu\tau)$, $Br(\tau\to\mu\gamma)$, muon g-2 and the other
experimental constraints as mentioned above.

In Fig. \ref{sigma}, we project the surviving samples on the planes
of the total width of 750 GeV singlet ($\Gamma_P$) versus $s_\theta$,
$Y_L$ versus $\Gamma_P$ and $Y_L$ versus $s_\theta$. From the left panel,
we can find that $\Gamma_P$ is very sensitive to $s_\theta$ since
the $P$ couplings to SM particles are relevant to $s_\theta$. The
$\Gamma_P$ value increases with the absolute value of $s_\theta$, and
reaches 35 GeV for $|s_\theta|=$0.3. No matter how large the
$\Gamma_P$ value is, the 750 GeV diphoton production rate
$R_{\gamma\gamma}$ can vary from 2 fb to 10 fb. The middle
panel shows that, with the increasing of the total width, $Y_L$ becomes
large enough to enhance $Br(P\to \gamma\gamma)$, and further
make $R_{\gamma\gamma}$ to be in the range of $2$ fb and $10$ fb.
For $\Gamma_P=$ 35 GeV, $R_{\gamma\gamma}>$ 4 fb requires $Y_L$ to
be larger than 30. The right panel shows that a large absolute value
of $s_\theta$ favors a large $Y_L$ since $s_\theta$ with a large
absolute value will enhance the total width of 750 GeV Higgs
sizably.

In Fig. \ref{width}, we project the surviving samples with
20 GeV$<\Gamma_P<$40 GeV on the planes of the widths of
the main decay modes of the 750 GeV Higgs singlet versus $\lambda_3$, $m_H$ and
$\lambda_7$. This figure shows that $P\to Ah$, $P\to HZ$, $P\to
AH$ and $P\to W^\pm H^\mp$ are the main decay modes, and the decay
$P\to hZ$ is insignificant due to the suppression of $s_\alpha$. The
decay $P\to Ah$ is sensitive to $\lambda_3$ and increases with
$\lambda_3$. The width of $P\to Ah$ can reach 20 GeV and dominate
over other decay modes for $\lambda_3=4\pi$. The decay $P\to HZ$
is sensitive to $m_H$ and decreases with the increasing of $m_H$.
The width of $P\to HZ$ can reach 10 GeV and be larger than those of
$P\to AH$ and $P\to W^\pm H^\mp$ for $m_H=200$ GeV. The decay $P\to
AH$ increases with $\lambda_7$ and can be larger than the width of
$P\to W^\pm H^\mp$ for $\lambda_7=10$. The width of $P\to W^\pm
H^\mp$ is in between 2 GeV and 4 GeV for $m_{H^\pm}=$ 500 GeV,
and not sensitive to $m_H$, $\lambda_3$ or $\lambda_7$. With the
increasing of $m_A$ and $m_{H^\pm}$, the decay $P\to hA$, $P\to HA$
and $P\to W^\pm H^\mp$ will be kinematically forbidden, which will
reduce the width of 750 GeV Higgs sizably.

\section{Conclusion}
To simultaneously accommodate the excesses of the 750 GeV diphoton, muon g-2
and $h\to\mu\tau$, we proposed an extension of 2HDM with vector-like
fermions and a CP-odd scalar singlet $P$, which is identified as the
750 GeV resonance. There is a mixing between the 750 GeV Higgs and the
CP-odd scalar $A$, which leads to the $P$ coupling to SM particles
and $A$ coupling to vector-like fermions. In the 2HDM the Higgs bosons
have tree-level LFV interactions with $\mu-\tau$, which can be
responsible for the excess of $h\to \mu\tau$ and
also give sizable contributions to the muon g-2.
The 750 GeV Higgs can decay into
$P\to Ah$, $P\to HZ$, $P\to AH$ and $P\to W^\pm H^\mp$,
and its total width is sensitive to $s_\theta$ and can reach 35 GeV
for $|s_\theta|=$ 0.3. Since the 750 GeV Higgs has a large width, the
vector-like leptons are required to enhance $Br(P\to
\gamma\gamma)$ to obtain $R_{\gamma\gamma}>$ 2 fb.
Meanwhile, such vector-like leptons will give sizable
contributions to $Br(\tau\to\mu\gamma)$ due to the mixing of $P$ and
$A$. Therefore, the Higgs couplings to $\bar{\tau}\tau$ are required
to be properly large to cancel the contributions of vector-like
leptons to $Br(\tau\to\mu\gamma)$. Considering the current
constraints of the LHC data, precision electroweak data and $Br(\tau\to\mu\gamma)$, we scanned over
the parameter space and found that such an extension can simultaneously
 explain the excesses of the 750 GeV diphoton, muon g-2 and $h\to\mu\tau$.

\section*{Acknowledgment}
This work has been supported in
part by the National
Natural Science Foundation of China under grant Nos. 11575152, 11305049, 11275057,
11405047, 11275245, 10821504 and 11135003,by the Spanish Government and ERDF funds
from the EU Commission [Grant No. FPA2011-23778],  by the Spanish {\it Centro de Excelencia
Severo Ochoa} Programme [Grant SEV-2014-0398], by the CAS Center for Excellence in
Particle Physics (CCEPP)

\end{document}